\documentclass[fleqn,usenatbib]{mnras}

\usepackage{newtxtext,newtxmath}
\usepackage[T1]{fontenc}
\usepackage{ae,aecompl}

\usepackage{graphicx}	% Including figure files
\usepackage{amsmath}	% Advanced maths commands
\title[Millisecond pulsar spectra]{On the radio spectra of Galactic millisecond pulsars}
\author[Aggarwal \& Lorimer]{
Kshitij Aggarwal$^{1, 2}$
and
D.~R. Lorimer$^{1, 2}$\thanks{E-mail: Duncan.Lorimer@mail.wvu.edu}
\\
$^{1}$West Virginia University, Department of Physics and Astronomy, P. O. Box 6315, Morgantown 26506, WV, USA\\
$^{2}$Center for Gravitational Waves and Cosmology, West Virginia University, Chestnut Ridge Research Building, Morgantown 26506, WV, USA\\
}

\date{}
\hypersetup{draft}

\pubyear{2021}

\begin{document}
\label{firstpage}
\pagerange{\pageref{firstpage}--\pageref{lastpage}}
\maketitle

\begin{abstract}
With recent advances in the sensitivity of radio surveys of the Galactic disk, the number of millisecond pulsars (MSPs) has increased substantially in recent years such that it is now possible to study their demographic properties in more detail than in the past. We investigate what can be learned about the radio spectra of the MSP population. Using a sample of 179 MSPs detected in eleven surveys carried out at radio frequencies in the range 0.135--6.6~GHz, we carry out detailed modeling of MSP radio spectral behaviour in this range. Employing Markov Chain Monte Carlo simulations to explore a multi-dimensional parameter space, and accurately accounting for observational selection effects, we find strong evidence in favour of the MSP population having a two-component power-law spectral model scaling with frequency, $\nu$.
Specifically, we find that MSP flux density spectra are approximately independent of frequency below 320~MHz, 
and proportional to $\nu^{-1.5}$ at higher frequencies. This parameterization performs significantly better than single power-law models which over predict the number of MSPs seen in low-frequency (100--200~MHz) surveys. We compared our results with earlier work, and current understanding of the normal pulsar population, and use our model to make predictions for MSP yields in upcoming surveys. We demonstrate that the observed sample of MSPs could triple in the coming decade.
\end{abstract}


%%%%%%%%%%%%%%%%%%%%%%%%%%%%%%%%%%%%%%%%%%%%%%%%%%

%%%%%%%%%%%%%%%%% BODY OF PAPER %%%%%%%%%%%%%%%%%%

\section{Introduction}
\label{sec:intro}
It has long been known \citep[see, e.g.,][]{sieber1973} that radio pulsars follow a flux density--frequency ($S_{\nu}$-$\nu$) dependence
in which the observed flux drops sharply with frequency. This behaviour can often be well approximated by a power-law:
\begin{equation}
    S_\nu \propto \nu^\alpha,
\end{equation}
where the parameter $\alpha$ is known as the spectral index. Commonly quoted mean values of $\alpha$  for the canonical pulsar\footnote{Canonical pulsars are sometimes referred to in the literature as ``normal'', ``slow'' or ``non-recycled'' pulsars.} (CP) population is --1.6 \citep{lorimer1995} or --1.8 \citep{maron2000}. More recently, when studying the Galactic population of CPs, \citet{bates2013} found them to be consistent with a Gaussian distribution having a mean of --1.4 and unit standard deviation.  For some pulsars, a single power-law model does not match the observations. In such cases, a broken power law with two spectral indices (a high frequency and low frequency spectral index) have been used \citep{sieber1973, lorimer1995, maron2000, jankowski2018}. \citet{Bilous2016, Bilous2020} even found evidence for a power law with more than one break for some CPs. 

Millisecond pulsars (MSPs) are the faster rotating cousins of CPs, and are characterized by their spin periods ($P$) and period derivatives ($\dot{P}$) which are a few orders of magnitude lower than the normal pulsars ($P\sim$ 1--20~ms, $\dot{P} \sim 10^{-19}\text{ss}^{-1}$ for MSPs and $P\sim$~1~s, $\dot{P} \sim 10^{-15}\text{ss}^{-1}$ for CPs). This results in the inferred values of surface magnetic fields for MSPs being three to four orders of magnitude lower than CPs \citep[for a recent review, see][]{Bhattacharyya2021}. 

Due to difficulties in their detection, the MSP population is observationally much less numerous than the CPs. As a result, meaningful constraints on MSP spectra have necessarily been quite limited. In an early study, \citet{foster1991} determined spectral indices of four MSPs, and found that this small group had a steeper spectra than for CPs. \citet{toscano1998} came to a similar conclusion using 19 MSPs, and obtained a mean value of $\alpha$ to be --1.9. However, in another study from this era,  \citet{kramer1998} using observations of MSPs with Effelsburg telescope argued that the spectra of MSPs and CPs are not significantly different. 

\citet{kuzmin2000, kuzmin2001} used observations of MSPs and concluded that MSP spectra showed no low frequency spectral turnover, unlike CPs. On the other hand, using synthesis imaging observations, \citet{kuniyoshi2015} found 
that 10 out of 39 MSPs observed below 100~MHz show signs of a low-frequency turnover, i.e.~a flatter spectrum at low frequencies compared to those at higher values. \citet{bassa2017, bassa2018} discovered three very steep spectrum MSPs from the location of unidentified {\it Fermi} gamma-ray sources, one of which was the second shortest period MSP currently known. \citet{kondratiev16} carried out low frequency flux measurements of 75 MSPs, and did not find much evidence for a spectral turnover. In another study, \citet{jankowski2018} found that spectra of a few MSPs could not be described by a simple power law.  Very recently, \citet{2021SCPMA..6429562W} discovered a faint MSP, J0318+0253, which shows evidence for a turnover in the radio spectrum at frequencies around 300~MHz.

As can be inferred from the above survey of the literature, while a significant body of information on MSP spectra now exists, like the CPs, there is quite a variation in spectral behaviour for individual objects and no clear consensus on any trends for the population. We attempt to address this situation by investigating the spectral indices of MSPs as population. Following a similar approach to the method of \citet{bates2013}, we simulate a pulsar population in the Milky Way whose underlying properties are well understood and employ detailed models of pulsar surveys to quantify observational selection effects. We use the relative yields of the various surveys to constrain the spectral behaviour for the MSP population as a whole. The rest of this paper is organised as follows. In \S2 we discuss the details of the population modeling, followed by the description of the simulations in \S3. \S4 and \S5 present the results and discussion of our analysis, followed by concluding remarks in \S6. 

\section{Methods}
\subsection{Population synthesis of pulsars}

Due to significant observational selection effects, it is well known that the observed sample of CPs in the Galactic disk, which currently exceeds 3000, is a small fraction of the total population \citep[estimated to exceed over $10^5$ pulsars; see, e.g.,][]{2006faucher}. In spite of this difficulty, because the observational selection effects are well understood, we can use Monte Carlo techniques to simulate the population and detection of pulsars in the Galaxy to understand the properties of the entire Galactic population and make predictions for future surveys. 

Generally, two strategies \citep[for a review, see, e.g.,][]{lorimer2019} are used for population synthesis of pulsars. In the first so-called ``snapshot'' approach, no assumptions are made concerning the prior evolution of pulsars. A pulsar population is generated using various distribution functions which are generally informed by previous studies. In the second method, sometimes known as the ``evolve'' approach, model pulsars are given initial birth parameters, and allowed to evolve forward in time using models of pulsar spin-down and the Galactic gravitational potential. In both cases, using accurate models of various large-scale pulsar surveys, mock samples of potentially detectable pulsars are compared to the actual sample of observed pulsars so as to allow detailed investigations of the input assumptions about the underlying pulsar population(s) and spin-down models.

\subsection{MSP sample and surveys}
\label{sec:survey_sample}
\begin{figure*}
	\includegraphics{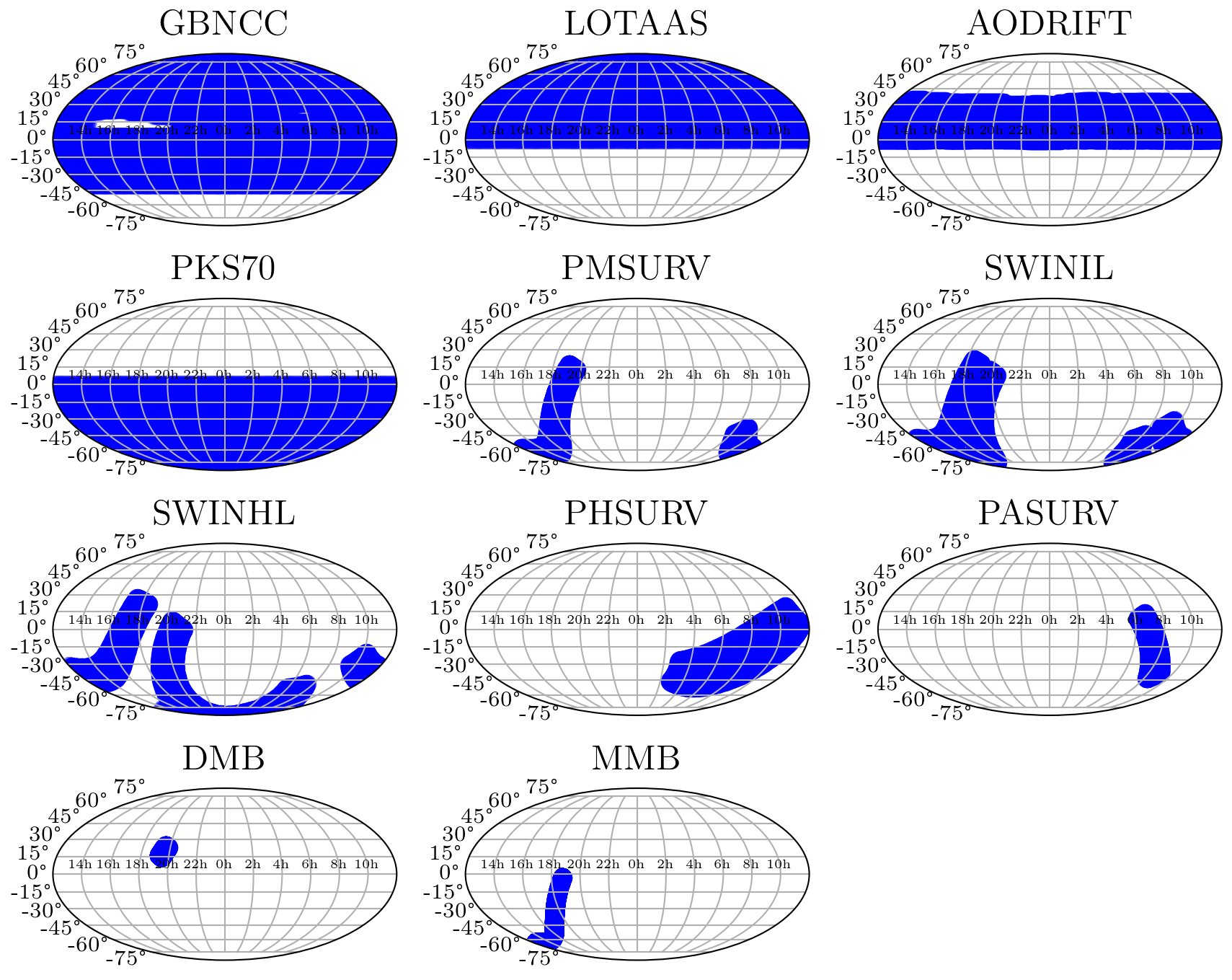}
    \caption{Hammer-Aitoff projections in celestial coordinates showing the approximate sky locations covered by different surveys used in this study. For a list of surveys used see Table~\ref{tab:survey_sample}.}
    \label{fig:pointings}
\end{figure*}

For this work, we define an MSP as a pulsar with spin period less than 20~ms. For the purposes of our population analysis, we used 11 surveys, given in Table~\ref{tab:survey_sample}. The total number of MSPs detected (i.e., discovered and re-detected) in this sample of surveys is 179. We used these surveys since these have already been extensively searched for MSPs using a variety of techniques, and therefore their number of detections should be reliable. Moreover, the surveys are at different frequency ranges which is necessary for confident estimates of spectral index. The sensitivity of these surveys is also well understood, therefore we can generate accurate survey models to simulate them in our simulations. These surveys also cover different and overlapping regions of the sky (see Fig.~\ref{fig:pointings}). For ongoing surveys like GBNCC, AODRIFT and LOTAAS, we only use the pointings that have been processed and have yielded the numbers reported in Table~\ref{tab:survey_sample}. We note that there are some surveys \citep[specifically HTRU and PALFA,][]{2010MNRAS.409..619K,2015ApJ...812...81L} which we have not considered explicitly here which were carried out at 1.4~GHz. We have omitted these for simplicity here, as they do not affect our conclusions about the spectral behaviour below 500~MHz. However, we do make use of them as a check of our predictions in \S4.2.

\begin{table}
\caption{Summary of MSPs detected by the surveys used in this analysis. From left to right, we list the survey mnemonic, telescope used, number of MSPs detected ($N_{\rm MSP}$) and the primary reference detailing the observational parameters and strategy.}
\centering
\label{tab:survey_sample}
\begin{tabular}{llcr}
    \hline
    Survey & Telescope & {$N_\mathrm{MSP}$} & Reference\\
    \hline
    DMB & Parkes & 2 & \citet{lorimer2013}\\
    LOTAAS & LOFAR & 10 & \citet{Sanidas2019}\\
    PKS70 & Parkes & 19 & \citet{bates2014}\\
    PHSURV & Parkes & 5 & \citet{burgay2006}\\
    MMB & Parkes & 0 & \citet{bates2014}\\
    GBNCC & GBT & 61 & \citet{mcewen2019}\\
    PASURV & Parkes & 1 & \citet{burgay2013}\\
    AODRIFT & Arecibo & 33 & \citet{Deneva2016}\\
    PMSURV & Parkes & 28 & \citet{lorimer2015}\\
    SWINHL & Parkes & 8 & \citet{edwards2001}\\
    SWINIL & Parkes & 12 & \citet{jacoby2007}\\
    \hline
\end{tabular}
\end{table}

\subsection{Population model parameters}

We used the pulsar population synthesis package \texttt{PsrPopPy} \citep{bates2014} to carry out all simulations reported in this paper. \texttt{PsrPopPy} has modules to carry out both the snapshot and evolve methods mentioned above. Although we investigated both approaches during the course of this study, to minimize the number of input assumptions, the final results presented in this paper made use of the snapshot approach to explore the dependence of spectral index of MSPs on survey yields. Throughout this work, we will refer to a family of six model MSP populations as A, B, C, D, E and F. All of the models use a log-normal luminosity function \citep[see, e.g.,][]{2006faucher} which introduces two parameters: the mean of the base-10 logarithm of the luminosity ($\langle\log_{10}L\rangle$) and a standard deviation in the same quantity ($\sigma_{\log_{10}L}$). In both cases, $L$ is expressed in units of mJy~kpc$^2$ and is defined to be at a reference frequency $\nu_{\rm ref}=1.4$~GHz. This choice of $\nu_{\rm ref}$ is appropriate since it reflects an intermediate point of the spectral range considered and the majority of MSP surveys used in our analysis were carried out at 1.4~GHz.
The MSPs are distributed spatially in a model galactic disk with a radial dependence found by \citep{lorimer2006} and using an exponential function to distribute the pulsars above and below the galactic plane with a scale height $z_0$. Assuming the Sun to be at (0.0,8.5,0.0)~kpc in a Cartesian coordinate system, we can then compute the distance to each model pulsar, $d$ and its corresponding flux density as seen from Earth when observed at the reference frequency, $S_{\rm ref}=L/d^2$. Strictly speaking, since this inverse square law scaling drops any geometrical factors, the luminosities used here are ``pseudoluminosities'' \citep[see, e.g.,][]{2002ASPC..278..227C}. However, since this definition is in common use throughout the pulsar literature, we refer to them henceforth as simply ``luminosities''.

To compute the flux density at different frequencies, we adopt power-law spectral models. In our simplest models (A--C), we choose spectral indices of each synthetic pulsar  from a Gaussian distribution with mean $\mu_{\alpha}$ and standard deviation $\sigma_{\alpha}$. Given a spectral index sampled from this distribution, for a survey carried out at observing frequency $\nu$, the flux density of each model MSP,
\begin{equation}
    S_{\nu} = S_{\nu_{\rm ref}} \left(
    \frac{\nu}{\nu_{\rm ref}}
    \right)^{\alpha}.
\end{equation}
For models D--F, we invoke a broken
power law parameterization of the flux density spectrum.
This results in two independent Gaussian
distributions with different means which are constructed such that the spectra intersect at a break frequency, $\nu_{\rm break}$.

All models adopt the log-normal MSP period distribution favoured by \citet{lorimer2015} and use a fixed pulse duty cycle of 20\% to compute the intrinsic pulse width. We assume for simplicity that there is no intrinsic pulse width evolution with frequency. For a detailed description of the modeling procedure employed in \texttt{PsrPopPy}, the reader is referred to \citet{bates2014}.

\subsection{Markov Chain Monte Carlo approach}
\label{sec:mcmc}
To obtain robust parameter estimations for each of our five models,
we use Markov Chain Monte Carlo (MCMC) simulations to efficiently explore the multi-dimensional parameter space and obtain the posterior distributions of the parameters introduced in the previous section. We used a pure python implementation of affine-invariant MCMC ensemble sampler, \texttt{emcee} \citep{goodman2010, foreman2013}. We used uniform priors and the parameter space was restricted between --2 and 0.0 for the power law index ($\mu_{\alpha})$ and 0--1 for the standard deviation ($\sigma_{\alpha}$). Priors used on other parameters are mentioned below. The procedure consists of the following steps: 
\begin{enumerate}
    \item given a set of spectral index parameters ($\mu_{\alpha}$ and $\sigma_{\alpha}$), generate a normal distribution of spectral indices;
    \item create a snapshot MSP population, sampling the properties  from the various distributions;
    \item generate MSPs until the total number of detections from all 11 surveys match the observed number of detections;
    \item evaluate  yields of individual surveys, and compare to the actual number of MSPs detected in those surveys. 
\end{enumerate}
The above steps are repeated multiple times. Each MCMC initially samples the values from the prior distribution of $\mu_{\alpha}$ and $\sigma_{\alpha}$, following which  the joint likelihood is used to estimate the subsequent trial parameters values. As mentioned previously, we used 11 surveys in our analysis, both to generate the observable population of model MSPs, and to compare the relative survey yields.     

We model the individual survey likelihoods using a Poisson distribution, and estimate the probability of detecting $N$ MSPs in a survey, given a simulation survey yield of $n$. As a result, the likelihood of finding $N$ events,
\begin{equation}
    {\cal L}(N|n)=\frac{n^{N} e^{-n}}{N!}.
\end{equation}
To compute the joint likelihood over all surveys considered requires the multiplication of small numerical values. In practice, the best way to perform such calculations is to 
calculate $\ln{\cal L}$ for all the surveys, and then add them together to obtain the joint log likelihood. 

\section{Results}
\label{sec:results}

\begin{figure}
	\includegraphics[width=0.4\textwidth]{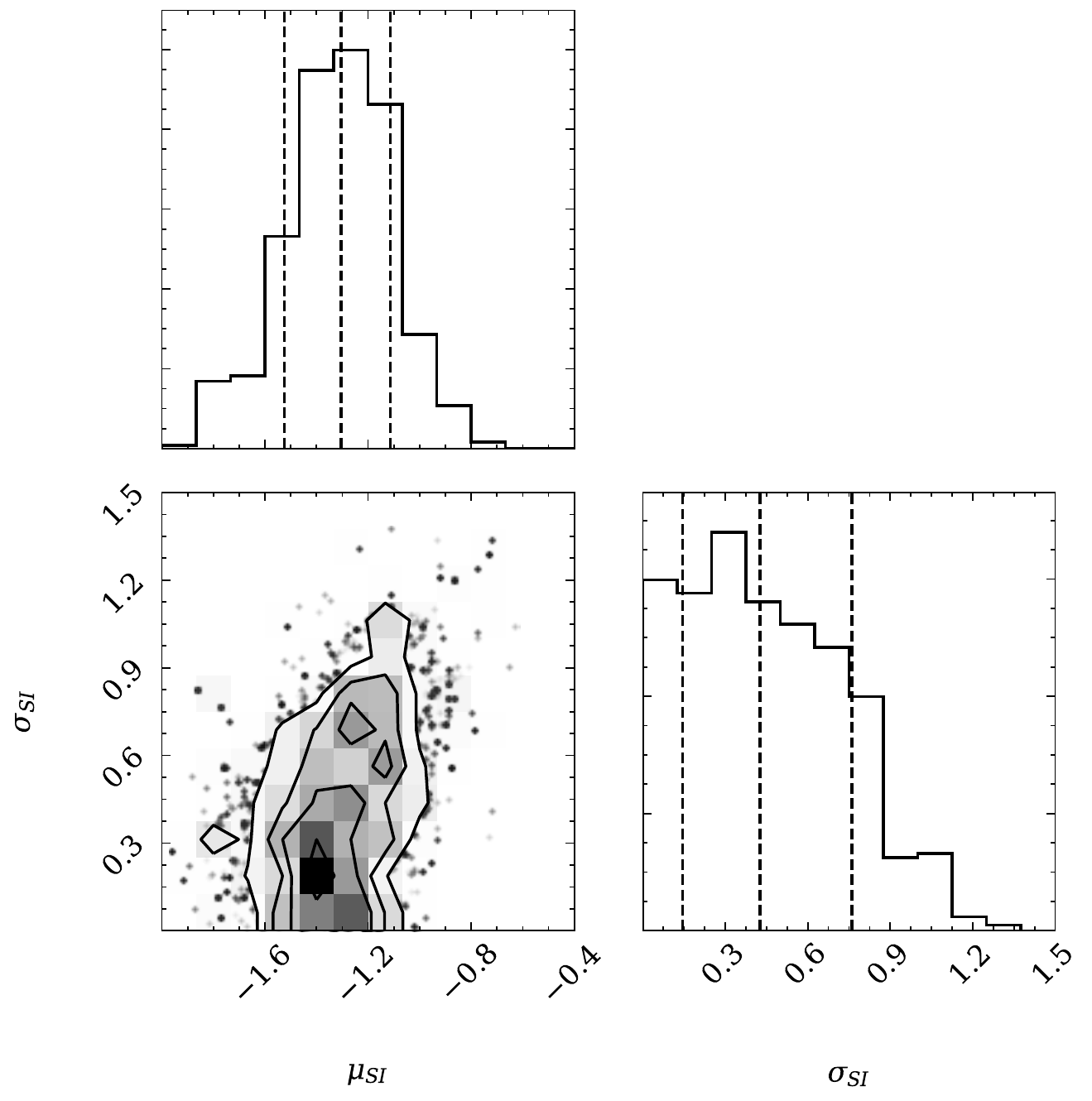}
\caption{Posterior distributions of $\mu_{\alpha}$ and $\sigma_{\alpha}$ for model A. The contours in the 2D histogram plot are at 1, 2 and 3~$\sigma$ levels.}
    \label{fig:si}
\end{figure}

Starting with our simplest model, model A, we show the posterior distributions of the mean and standard deviation of spectral index in Fig.~\ref{fig:si}. The median values of the two parameters are $\mu_{\alpha}=-1.3 \pm 0.2$ and $\sigma_{\alpha}=0.43^{+0.33}_{-0.28}$ respectively\footnote{All  errors reported in this paper are 1$\sigma$ confidence intervals.}. The posterior distribution of $\sigma_{\alpha}$ deviates significantly from a Gaussian distribution, and extends towards lower values. In the subsections below, we describe how we explored the various underlying assumptions in model A by showing the results of the other models (B--F).

\subsection{Dependence on vertical scale height}
We next tried to estimate the scale height distribution of the MSPs along with the spectral index. We did this because the detectability of pulsars is sensitive to their scale height as it influences the received flux from the pulsar. In \texttt{PsrPopPy}, model pulsars are drawn from a two-sided exponential \citep{bates2014} probability density function $p(z)$, characterized by a scale height $z_0$. When normalized, this distribution can be written in differential form as
\begin{equation}
    p(z) \, {\rm d}z = \frac{1}{2z_0}\exp\left(\frac{-|z|}{z_0}\right) \, {\rm d}z.
\end{equation}
Therefore, changing the scale height factor ($z_0$) would change the distribution (and distances) of pulsars in the Galaxy, that will in-turn influence their detectability. We therefore tried to constrain the scale height factor along with the spectral index parameters in our MCMC. For the scale height, we used uniform priors of range 0.1 to 1.5~kpc. We fixed $\sigma_{\alpha}$ to be 0.7 and used the same surveys and priors for $\mu_{\alpha}$ as described previously (see Section~\ref{sec:mcmc}).

\begin{figure}
	\includegraphics[width=0.4\textwidth]{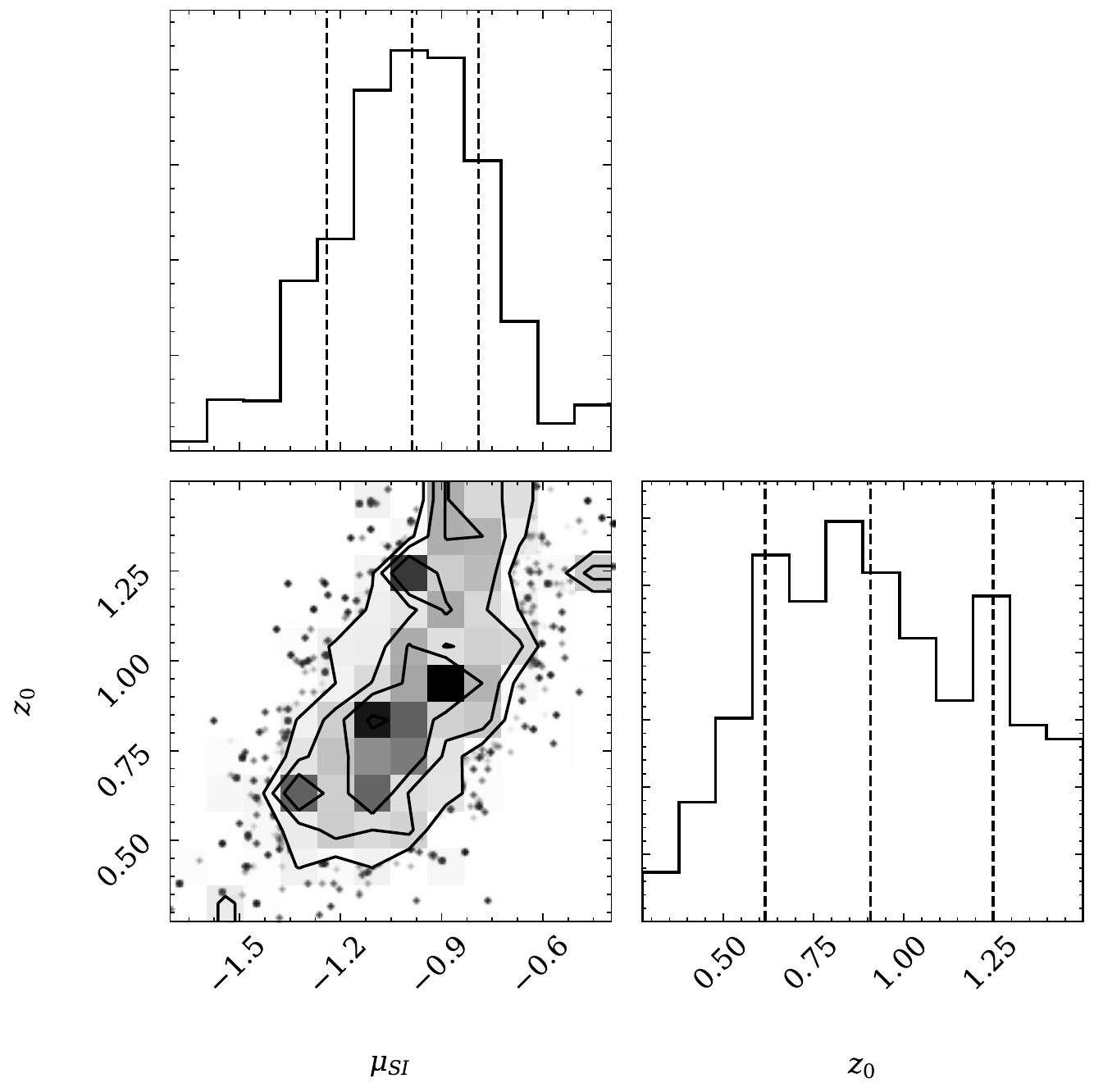}
    \caption{Posterior distributions of $\mu_{\alpha}$ and $z_0$ for model B. The contours in the 2D histogram plot are at 1, 2 and 3~$\sigma$ levels. $\sigma_{\alpha}$ was set at 0.7 for this simulation. }
    \label{fig:si_z}
\end{figure}

The results of this simulation are given in Fig.~\ref{fig:si_z} and referred to as model B. The converged values of the two parameters in this case are: $\mu_{\alpha}=-0.99^{+0.20}_{-0.25}$, and $z_{0}=0.91^{+0.34}_{-0.29}$ respectively. The posterior distribution of $z_0$ deviates from a Gaussian distribution, with a flat posterior between 0.6--1.25~kpc. While this implies that large values of scale height are preferred in the simulations,
we found that our results are not significantly impacted by the choice of $z_0$. In previous work
\citep[see, e.g.,][]{1995MNRAS.274..300L} it has been found that
scale heights as low 
as 500~pc provide an
adequate description of the observed MSP population. This constraint is applied in models A, D, E, and F and is found to work very well, particularly for models D--F.

\subsection{Dependence on radio luminosity distribution}
Next, we included the luminosity distribution of the MSPs, along with the scale height and spectral index distributions, to our MCMC framework. The detectability of the pulsars is dependent on the luminosity of the pulsars, and therefore we tried to estimate the mean of the log normal distribution ($\langle\log_{10}L\rangle$) of luminosity. We fixed $\sigma(\log_{10}$L) at 0.9, used uniform priors of range --2 to 0 for $\langle\log_{10}L\rangle$ and same methods as that in the previous subsection.

\begin{figure}
	\includegraphics[width=0.45\textwidth]{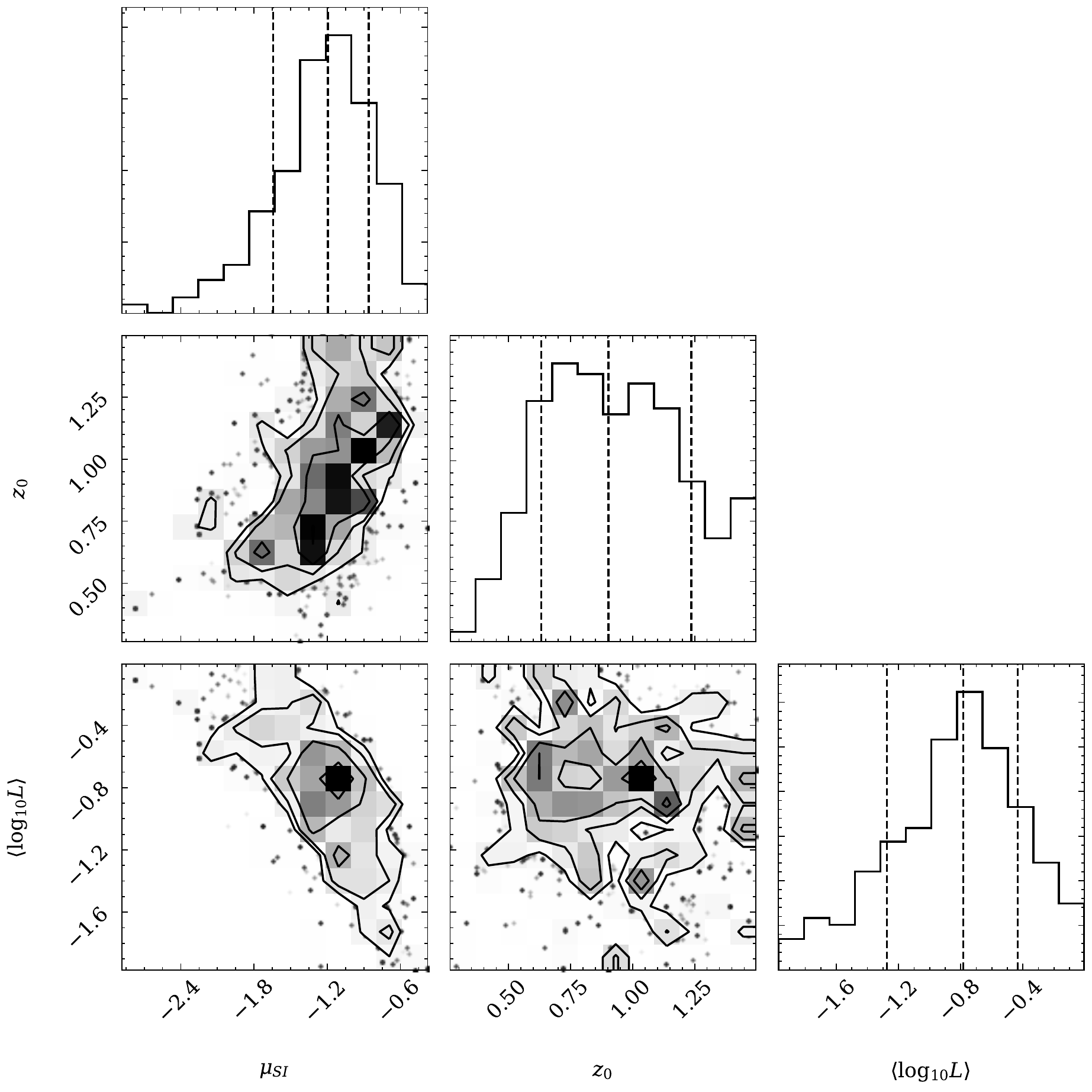}
    \caption{Posterior distributions of $\mu_{\alpha}$, $z_0$ and $\langle\log_{10}L\rangle$ for model C. The contours in the 2D histogram plot are at 1, 2 and 3~$\sigma$ levels. $\sigma_{\alpha}$ and $\sigma(\log_{10}$L) were set at 0.7 and 0.9 for this simulation.}
    \label{fig:si_z_l}
\end{figure}

Model C encapsulates these parameters, and the results of this simulation are given in Fig.~\ref{fig:si_z_l}. The converged values of the three parameters in this case are: $\mu_{\alpha}=-1.19^{+0.33}_{-0.45}$, $z_{0}=0.90^{+0.33}_{-0.27}$ and $\langle\log_{10}L\rangle = -0.78^{+0.35}_{-0.49}$ respectively. The posterior distribution of $z_0$ again deviates from a Gaussian distribution, with a flat posterior between 0.6--1.25~kpc. 

\subsection{Two-component spectral models}

As mentioned in Section~\ref{sec:intro}, a low-frequency break (or a turnover) has been reported for many CPs and MSPs. We therefore incorporated this in our population simulation and tested the presence of a  break between two discrete power laws in the spectra of MSPs with our MCMC framework. We modeled the two indices of the MSP using two different Gaussian distributions (each with a mean and a standard deviation), and a break frequency. We tried the following approaches implemented in models D--F. This work was motivated by the fact that population models generated using the results of previous simulations overestimated the number of MSPs detected in the low-frequency surveys.

\begin{figure}
	\includegraphics[width=0.45\textwidth]{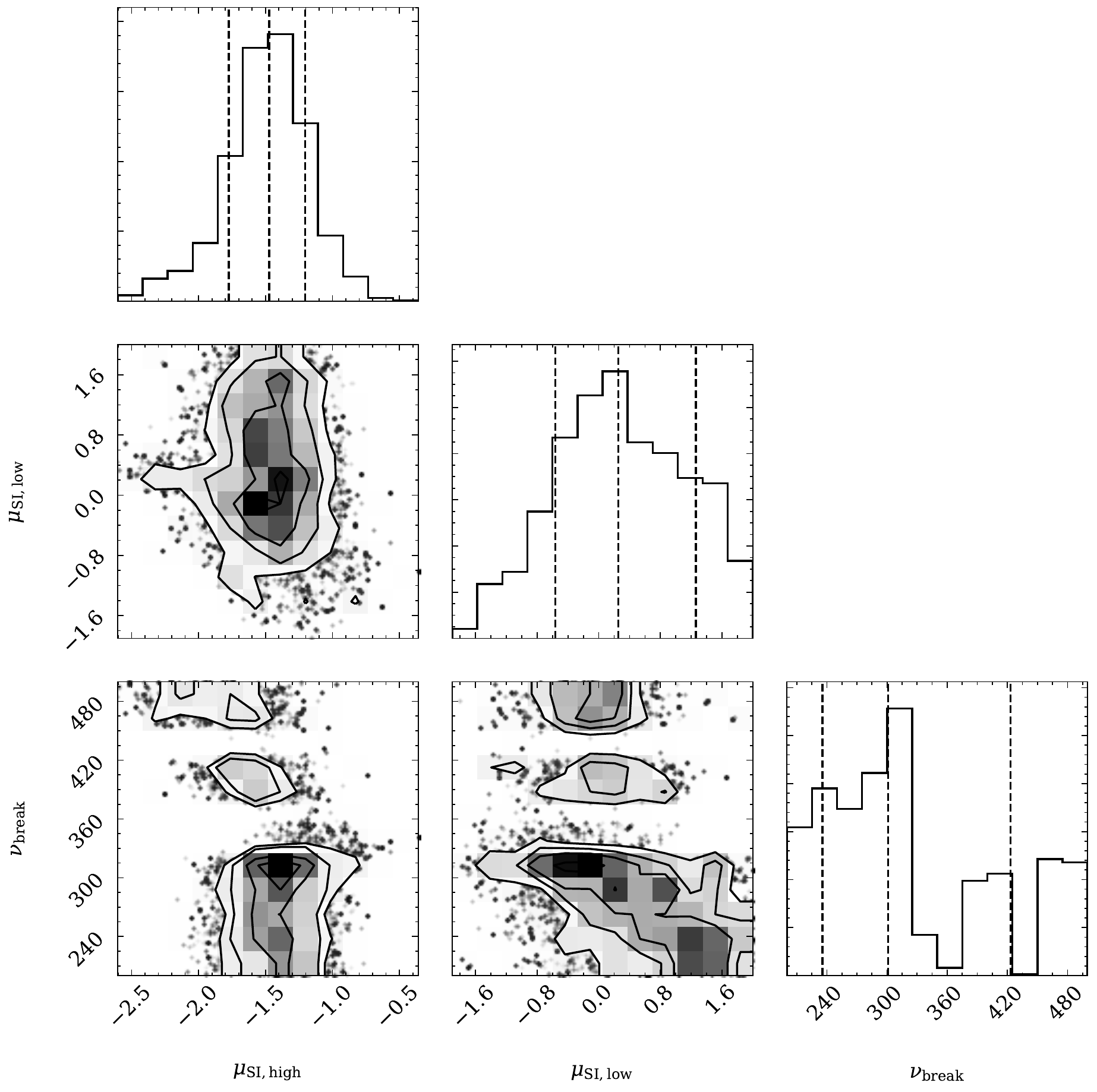}
    \caption{Posterior distributions of $\mu_{\alpha,{\rm low}}$, $\mu_{\alpha,{\rm high}}$ and $\nu_{\rm break}$ for model D. The contours in the 2D histogram plot are at 1, 2 and 3~$\sigma$ levels. $\sigma_{\alpha, {\rm high}}$ and $\sigma_{\alpha, {\rm low}}$ were both set at 0.7 for this simulation.}
    \label{fig:si_break}
\end{figure}

\begin{figure}
	\includegraphics[width=0.4\textwidth]{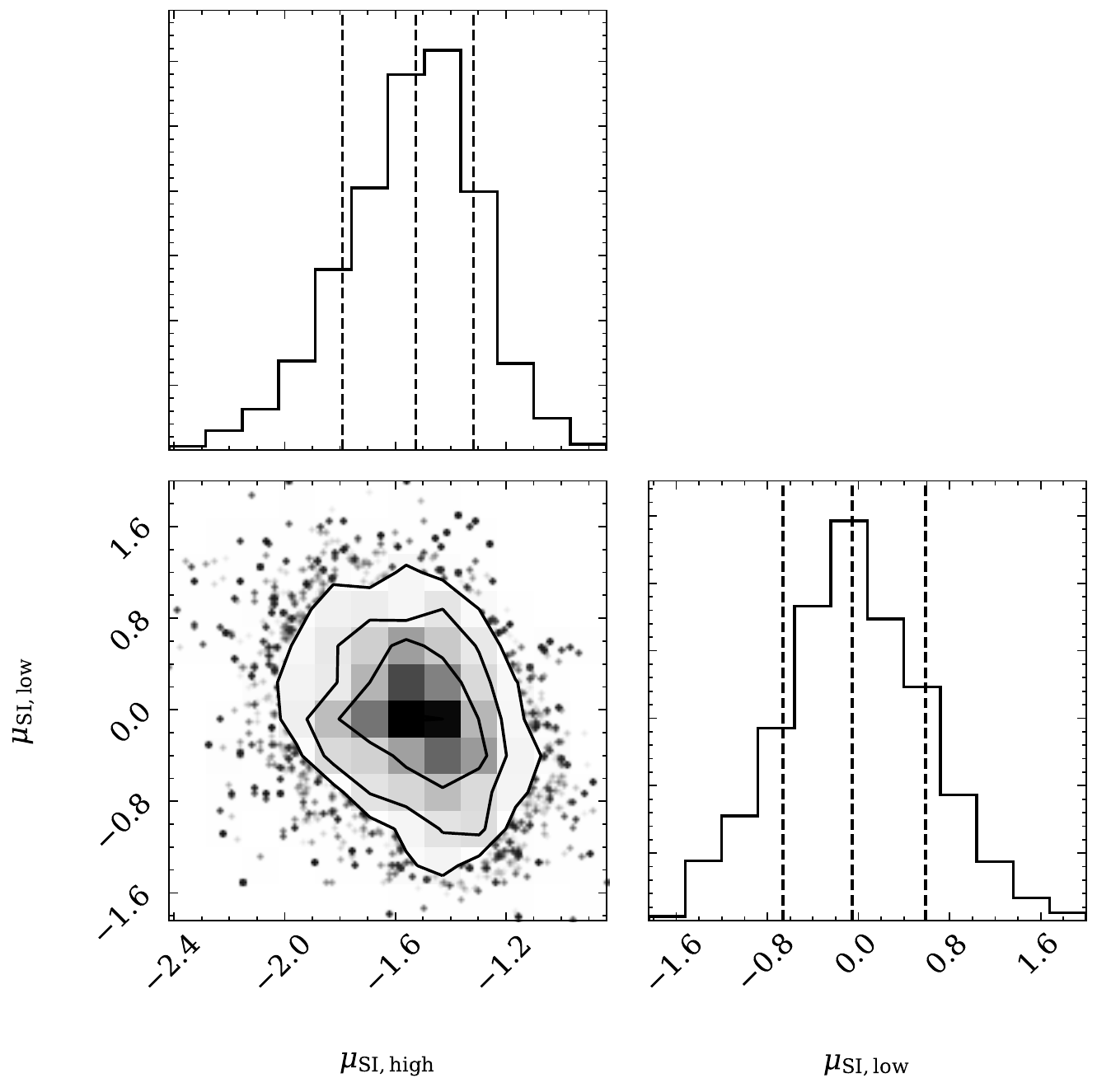}
    \caption{Posterior distributions of $\mu_{\alpha,{\rm low}}$ and $\mu_{\alpha,{\rm high}}$ for model E. The contours in the 2D histogram plot are at 1, 2 and 3~$\sigma$ levels. $\nu_{\rm break}$, $\sigma_{\alpha, {\rm high}}$ and $\sigma_{\alpha, {\rm low}}$ were set at 400~MHz, 0.7 and 0.7 for this simulation.}
    \label{fig:si_break_fixed_freq}
\end{figure}

In the first approach (model D) we fixed the standard deviations of the two indices to 0.7 and with uniform priors on the break frequency between range 200 to 500~MHz. Uniform priors between ranges: (--4, 2) and (--4, 0) were used for the mean of low frequency spectral index ($\mu_{\alpha,{\rm low}}$) and high frequency spectral index ($\mu_{\alpha,{\rm high}}$) respectively. This was done to accommodate not just a spectral break, but also a turnover at lower frequency. Finding a relatively large uncertainty on
the break frequency, we considered a second approach (model E) in which the break frequency was set to 400~MHz. The results of these two models are shown in Figs.~\ref{fig:si_break} and \ref{fig:si_break_fixed_freq}. The converged values of the parameters are: 
$\mu_{\alpha,{\rm low}}=-0.26^{+1.01}_{-0.82}$,  $\mu_{\alpha,{\rm high}}=-1.47^{+0.27}_{-0.30}$ and $\nu_{\rm break}=301^{+122}_{-65}$~MHz respectively for model D, and $\mu_{\alpha,{\rm low}}=-0.05^{+0.64}_{-0.61}$ and $\mu_{\alpha,{\rm high}}=-1.53^{+0.21}_{-0.27}$ for model E.
In both cases, the high frequency spectral index is steeper than the low frequency spectral index. 

\begin{figure}
	\includegraphics[width=0.4\textwidth]{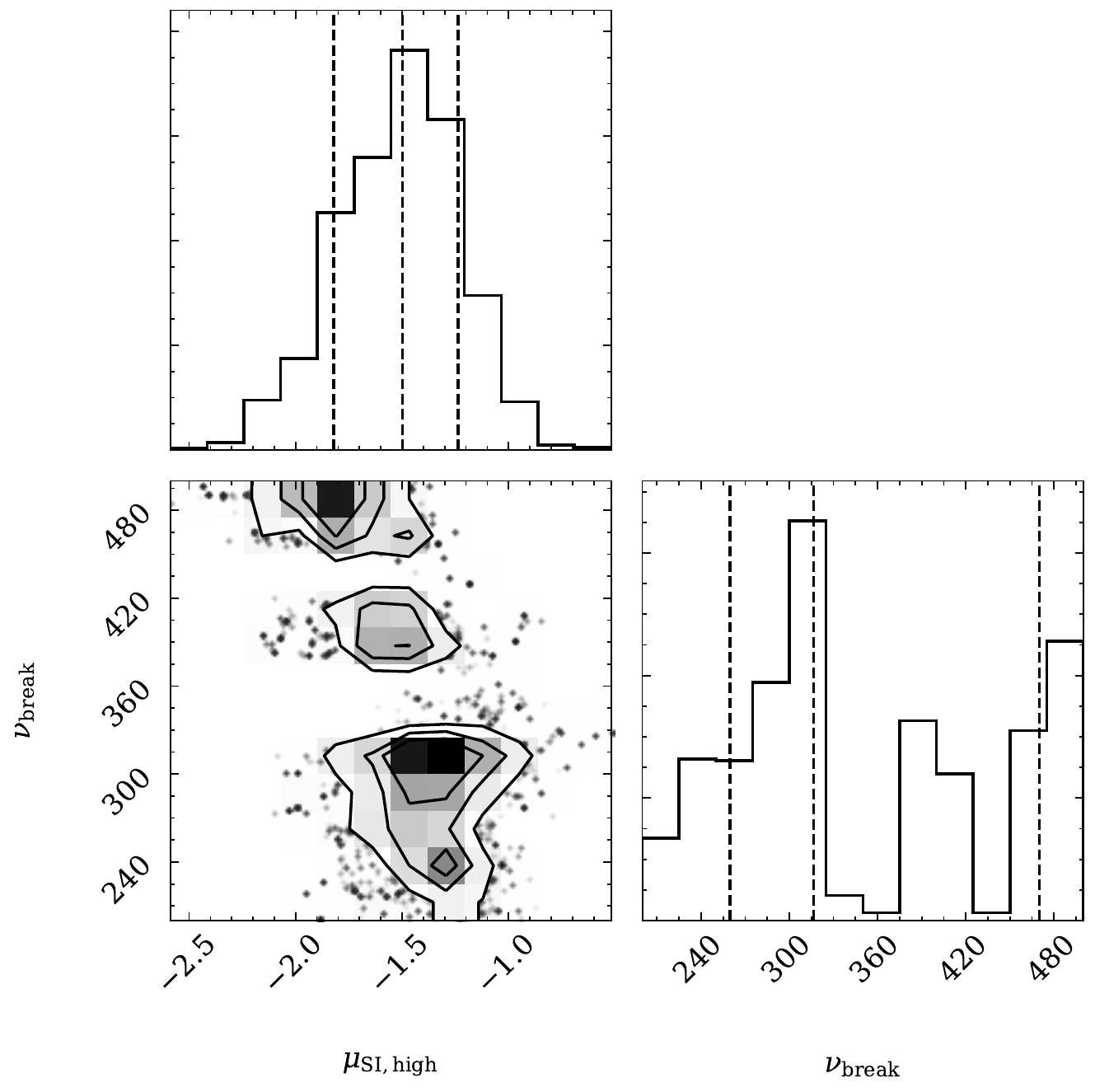}
    \caption{Posterior distributions of $\mu_{\alpha,{\rm high}}$ and $\nu_{\rm break}$ for model F. The contours in the 2D histogram plot are at 1, 2 and 3~$\sigma$ levels. $\mu_{\alpha,{\rm low}}$, $\sigma_{\alpha, {\rm low}}$ and $\sigma_{\alpha, {\rm high}}$ were set at 0.0, 0.7 and 0.7 for this simulation.}
    \label{fig:si_break_fixed_low_si}
\end{figure}

Because both models D and E suggested a relatively weak dependence on spectral behaviour with frequency in the lowest frequency component,
we considered a final model (F) in which the break frequency was allowed to vary, but  $\mu_{\alpha,{\rm low}}$ was constrained
to be zero (i.e.~no low-frequency spectral dependence). The converged values for this model are:
$\mu_{\alpha,{\rm high}}=-1.50^{+0.26}_{-0.32}$ and $\nu_{\rm break}=320^{+150}_{-60}$~MHz and is shown in Fig~\ref{fig:si_break_fixed_low_si}. 

\subsection{Comparing models} \label{sec:metric}

In the previous sections, we have discussed six models that we tested to constrain the pulsar population parameters, primarily the spectral index. The final converged parameters for each model are given in Table~\ref{tab:converged_parameters}. To compare the models, we use two complementary approaches.
In the first approach, from the likelihood values of each model considered, we computed its likelihood ratio ($\cal B$) relative to the model with the highest likelihood (model D in our case). The corresponding Bayes factor for model D versus each of the other models are the reciprocals of these ratios. Following \citet{jeffreys}, we deem models for which $0.3 < {\cal B} < 1$ to be indistinguishable from one another while ${\cal B} < 0.01$ are decisively ruled out in favor of models with higher values of ${\cal B}$. Models D--F are decisively favored over models A--C.

In the second approach, we obtain a numerical measure of how well each model reproduces the actual survey yields. For each model, we ran the ``snapshot'' simulations 100 times at the converged parameters, and then compared the survey yields in each case.  To compare the survey yields predicted by each model to those found in the actual surveys, we used a reduced $\chi^2$-style estimate which we refer to as the survey yield metric,  
\begin{equation}
    {\cal M} = \frac{1}{(s-1)}\sum_{i} \frac{(o_i - e_i)^2}{o_i},
\end{equation}
where, for a set of $s$ surveys, $o_i$ is the observed number of MSPs discovered in $i^{\rm th}$ survey, and $e_i$ is the estimated number of MSPs discovered, predicted by the simulations. We calculate the value of ${\cal M}$ for each of the hundred runs, for all the models. Models which score the lowest values of {$\cal M$} provide the best match to the observed survey yields. As anticipated from their likelihoods, models D--F perform consistently better than models A--C as measured by the survey yield metric. The complete set of survey yields and predictions for the observed sample and all six models,
with their corresponding {$\cal M$} and {$\cal B$} values are given in Table 4.

\begin{table*}
\caption{Summary of the parameter estimation analysis for all models presented in this paper (A--F). Values without errors were kept constant during the MCMC. From left to right, we list the model name, low frequency spectral index ($\mu_{\alpha,{\rm low}}$), high frequency spectral index ($\mu_{\alpha,{\rm high}}$), standard deviation in spectral index distribution(s) ($\sigma_{\alpha}$), scale height ($z_{0}$), mean log-luminosity ($\langle\log_{10}L\rangle$), standard deviation of log-luminosity  ($\sigma_L$), the likelihood ratio with respect to model D ($\cal B$) and the value of the survey yield metric ($\cal M$).}
\centering
\label{tab:converged_parameters}
\begin{tabular}{crrcccccrr}
\hline 
Model & \multicolumn{1}{c}{$\mu_{\alpha,{\rm low}}$} & \multicolumn{1}{c}{$\mu_{\alpha,{\rm high}}$} & $\sigma_{\alpha}$ & $\nu_{\rm break}$ & $z_{0}$ & $\langle\log_{10}L\rangle$ & $\sigma_L$  &  \multicolumn{1}{c}{$\cal B$} & \multicolumn{1}{c}{$\cal M$} \\
 & &  & & (MHz) & (pc) & &    &  & \\
\hline
A & -- &$-1.3 \pm 0.2$ & 0.43$^{+0.33}_{-0.28}$ & -- & 500 & --1.1 & 0.9&  0.002 & 3.6$^{+1.4}_{-1.2}$\\
B & -- &$-1.0 \pm 0.2$ & 0.7 & -- & 910$^{+340}_{-290}$ & --1.1 & 0.9 & 0.003 & 3.2$^{+1.6}_{-1.3}$\\
C & -- &--1.2$^{+0.3}_{-0.5}$ & 0.7 & -- & 900$^{+330}_{-270}$ & --0.78$^{+0.35}_{-0.49}$ & 0.9 & 0.002 & 3.0$^{+1.2}_{-1.1}$\\
D & 0.26$^{+1.01}_{-0.82}$ &--1.47$^{+0.27}_{-0.3}$  & 0.7 & 300$^{+120}_{-65}$ & 500 & --1.1 & 0.9 & 1 & 1.3$^{+0.7}_{-0.5}$\\
E & --0.05$^{+0.64}_{-0.61}$ &--1.53$^{+0.21}_{-0.27}$  & 0.7 & 400 & 500 & --1.1 & 0.9 & 0.29 & 1.8$^{+1.3}_{-0.6}$\\
F & 0.0 & --1.5$^{+0.26}_{-0.32}$ & 0.7 & 320$^{+150}_{-57}$ & 500 & --1.1 & 0.9 & 0.59 & 1.3$^{+0.8}_{-0.5}$\\
\hline
\end{tabular}
\end{table*}

\section{Discussion}
\label{sec:discussion}

Based on the results presented in the previous section,
we find strong evidence in favor of a two-component
population of MSP spectra (models D--F) which can
replicate the yields of surveys carried out at a variety of frequencies far better than models which invoke
a single power-law spectrum (models A--C). We discuss our findings in detail below.

\subsection{General remarks}

\begin{figure}
	\includegraphics[width=0.45\textwidth]{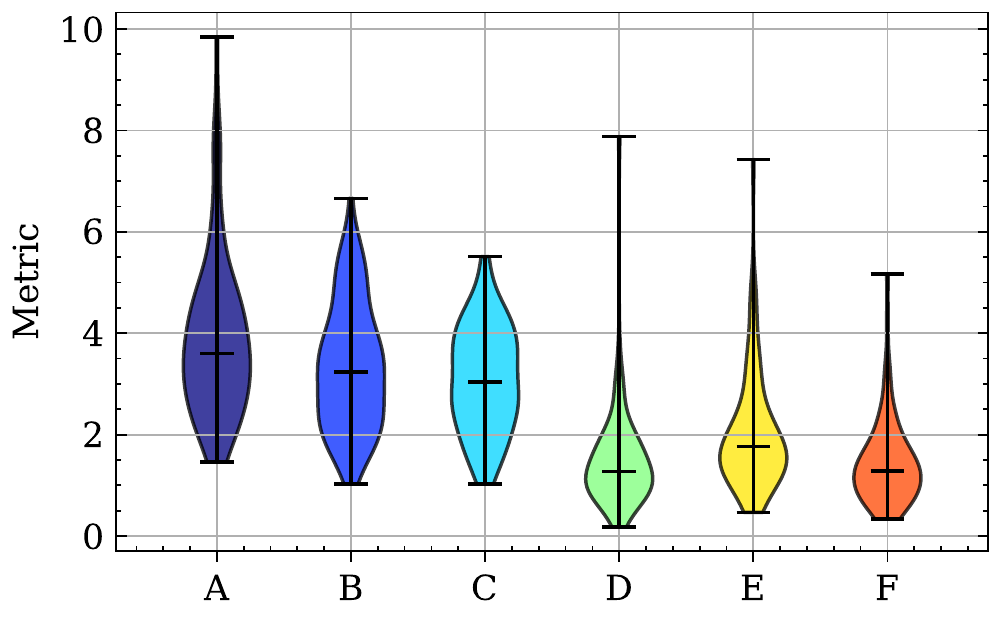}
    \caption{Violin plots of the survey yield metric for each model. Different colors represent different models. For each model, we show the distribution of metrics obtained on evaluating survey yields 100 times at the converged parameters listed in Table~\ref{tab:converged_parameters}.}
    \label{fig:metric_violin_plot}
\end{figure}

Tables~\ref{tab:converged_parameters} and \ref{tab:hundred_run_results} show the results of model comparison discussed in the previous section.  Table~\ref{tab:hundred_run_results} also shows the median survey yields obtained for each model, along with the 1$\sigma$ variance in the predictions. Fig.~\ref{fig:metric_violin_plot} shows the distribution of the survey yield metric (${\cal M}$) for all the models, along with the median values in each case. Fig.~\ref{fig:metric_violin_plot} and Table~\ref{tab:hundred_run_results} clearly show that value of the survey yield metric, ${\cal M}$, is the smallest for models D and F, while it is highest for model A. The spread in the distribution of the survey yield metric shows the robustness of the parameters across multiple runs. The distributions for model D--F lie at lowest values of the survey yield metric, while the distributions have a much larger spread for models A--C. Formally, based on its slightly higher likelihood and lowest value of the survey yield metric, model D represents our best characterization of the spectral index distribution.

Using Table~\ref{tab:hundred_run_results}, we can compare the exact survey yields of these three models of interest. Model B overestimates the yield for LOTAAS, while it underestimates that for AODRIFT, and slightly underestimates the yields for GBNCC and PKS70. In general, models that had the break in the spectral index (model D and E) result in best estimates for individual surveys. Of the models with a single power-law spectral index, model B performs best.

\begin{figure}
	\includegraphics[width=0.45\textwidth]{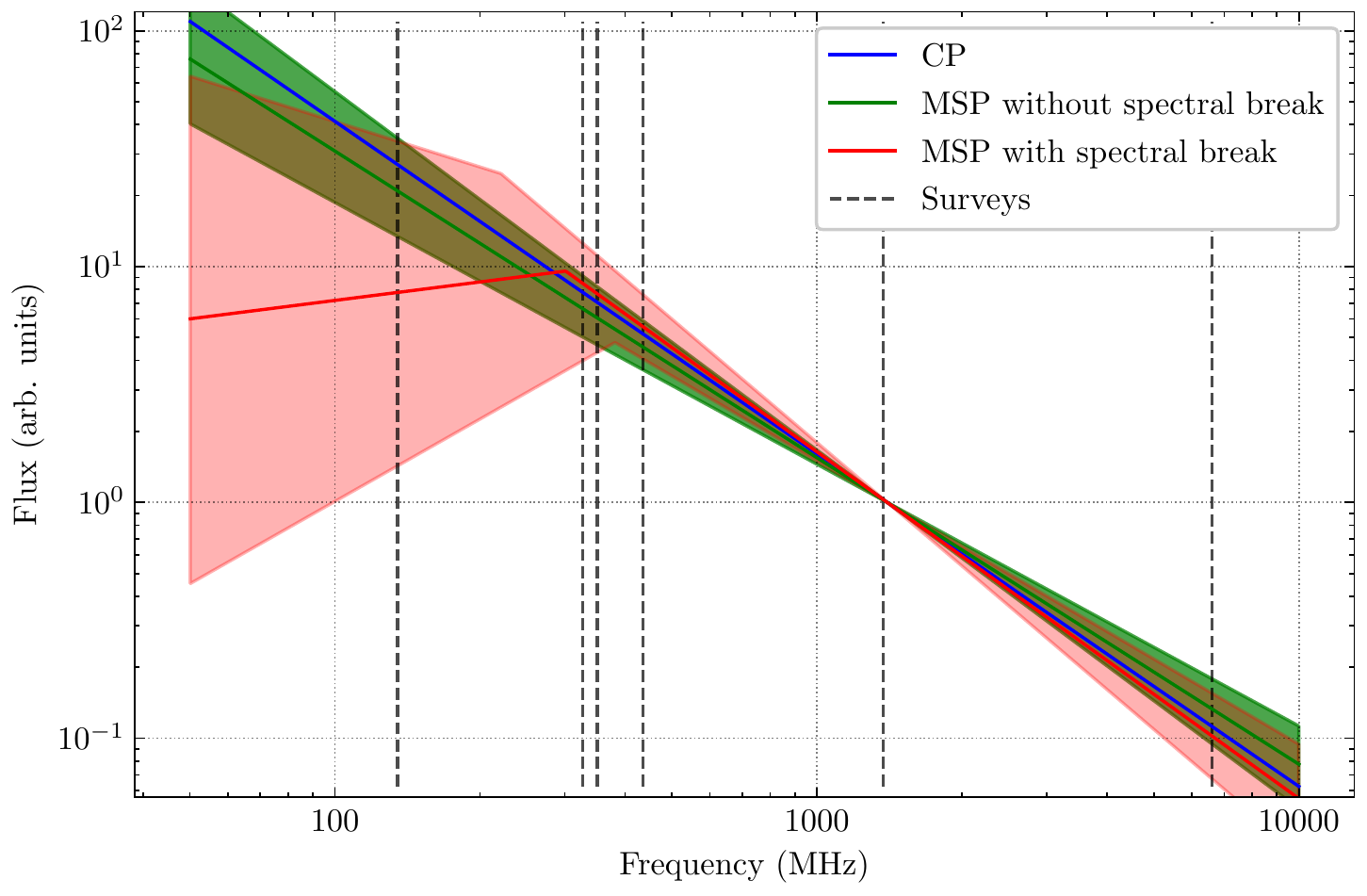}
    \caption{Visualizing the spectral index. Different colors represent different spectral indices: Blue represents the estimate by \citet{bates2013}, green represents model B and red shows the spectral index at model D. Shaded region represents the error region on the indices. Black vertical dotted lines show the frequency of surveys used in this work.}
    \label{fig:si_viz}
\end{figure}

If we assume that MSP spectra follow a single power law, then our results imply that LOTAAS should be able to find more MSPs ($\sim$10 more) from the already processed data. This can be either be explained by missed detections in LOTAAS (perhaps due to scintillation), or by invoking a low frequency break in the spectral index, with a flatter spectral index at lower frequencies (see Fig.~\ref{fig:si_viz}). Survey yields for models D and E (see Table~\ref{tab:hundred_run_results})  show that using a break in spectral index improves the predictions. These models have more accurate yields for all surveys, including the low frequency surveys. Table~\ref{tab:converged_parameters} shows that the low frequency spectral index for these models is much flatter (or even inverted) compared to the high frequency spectral index. 

\begin{table*}
\caption{Table showing the median number of MSPs discovered in the surveys (with 1~$\sigma$ variation) obtained by running the simulations 100 times at the converged parameter values obtained using MCMC. The bottom rows show the median metric value (${\cal M}$) with 1~$\sigma$ variation along with the likelihood ratio (${\cal B}$) relative to model D. Single power-law models (A, B and C) overestimate the detections for LOTAAS while they underestimate those for AODRIFT, GBNCC and PKS70. This indicates the presence of a spectral break around 300~MHz. Models with a break in the spectra (D, E and F) more accurately predict survey yields for all the surveys used in this analysis. Also listed is the implied number of potentially observable (i.e.~those sources beamed towards us) MSPs for each model.}
\centering
\label{tab:hundred_run_results}
\begin{tabular}{lcrrrrrrr}
    % \noalign{\smallskip} 
    \hline 
    % \noalign
    Survey & $\nu_{\rm obs}$ & $N_{\rm det}$ & \multicolumn{6}{c}{Model} \\
         &  (MHz) &   & A &  B & C & D & E & F\\
    \hline
    LOTAAS & 135 & 10 & 22$^{+4}_{-5}$ & 21$^{+4}_{-4}$ & 20$^{+4}_{-4}$ & 10$^{+3}_{-3}$ & 10$^{+4}_{-3}$ & 11$^{+3}_{-3}$\\
    AODRIFT & 327 & 33 & 22$^{+4}_{-4}$ & 25$^{+5}_{-4}$ & 24$^{+5}_{-4}$ & 38$^{+4}_{-6}$ & 23$^{+4}_{-4}$ & 36$^{+4}_{-6}$\\
    GBNCC & 350 & 61 & 58$^{+4}_{-6}$ & 59$^{+5}_{-4}$ & 62$^{+4}_{-6}$ & 64$^{+5}_{-6}$ & 65$^{+5}_{-5}$ & 67$^{+4}_{-7}$\\
    PKS70 & 436 & 19 & 16$^{+4}_{-3}$ & 16$^{+3}_{-3}$ & 18$^{+4}_{-3}$ & 18$^{+4}_{-4}$ & 22$^{+4}_{-3}$ & 18$^{+4}_{-4}$\\
    DMB & 1374 & 2 & 2$^{+1}_{-2}$ & 1$^{+1}_{-1}$ & 1$^{+1}_{-1}$ & 1$^{+2}_{-1}$ & 2$^{+1}_{-1}$ & 1$^{+2}_{-1}$\\
    PHSURV & 1374 & 5 & 4$^{+3}_{-2}$ & 4$^{+2}_{-2}$ & 3$^{+2}_{-1}$ & 4$^{+1}_{-2}$ & 4$^{+2}_{-2}$ & 3$^{+2}_{-1}$\\
    PASURV & 1374 & 1 & 3$^{+2}_{-2}$ & 2$^{+2}_{-1}$ & 2$^{+1}_{-1}$ & 2$^{+2}_{-1}$ & 3$^{+2}_{-2}$ & 2$^{+2}_{-1}$\\
    PMSURV & 1374 & 28 & 28$^{+5}_{-5}$ & 21$^{+5}_{-3}$ & 22$^{+5}_{-6}$ & 24$^{+4}_{-6}$ & 27$^{+7}_{-4}$ & 23$^{+4}_{-5}$\\
    SWINHL & 1374 & 8 & 10$^{+3}_{-3}$ & 13$^{+3}_{-4}$ & 13$^{+3}_{-4}$ & 8$^{+2}_{-2}$ & 9$^{+3}_{-3}$ & 7$^{+3}_{-2}$\\
    SWINIL & 1374 & 12 & 14$^{+3}_{-4}$ & 14$^{+3}_{-3}$ & 15$^{+3}_{-4}$ & 11$^{+3}_{-4}$ & 13$^{+2}_{-3}$ & 11$^{+3}_{-3}$\\
    MMB & 6600 & 0 & 0$^{+1}_{-0}$ & 1$^{+0}_{-1}$ & 0$^{+1}_{-0}$ & 0$^{+1}_{-0}$ & 0$^{+1}_{-0}$ & 0$^{+1}_{-0}$\\
    %  & Total & 179 & 3.6$^{+1.4}_{-1.19}$ & 3.24$^{+1.6}_{-1.26}$ & 3.04$^{+1.16}_{-1.11}$ & 1.28$^{+0.72}_{-0.46}$ & 1.77$^{+1.31}_{-0.57}$ \\
    % \hline
    \hline
    \multicolumn{2}{l}{Metric, ${\cal M}$}  & &3.6$^{+1.4}_{-1.2}$ & 3.2$^{+1.6}_{-1.3}$ & 3.0$^{+1.2}_{-1.1}$ & 1.3$^{+0.7}_{-0.5}$ & 1.8$^{+1.3}_{-0.6}$ & 1.3$^{+0.8}_{-0.5}$ \\
    \multicolumn{2}{l}{Likelihood ratio (relative to D), ${\cal B}$} & &
    0.002 & 0.003 & 0.002 & 1 & 0.29 & 0.59 \\
    \multicolumn{2}{l}{Total number of potentially observable MSPs ($\times 10^4$)} & &
    $2.8 \pm 0.3$ &
    $3.5 \pm 0.4$ & $1.6 \pm 0.2$ & $2.3 \pm 0.2$ & $2.7 \pm 0.3$ & $2.3 \pm 0.2$   \\
    \hline
    % \bottomrule
\end{tabular}
\end{table*}

\subsection{Caveats}

It is important to note some limitations to our analysis. As mentioned, we used the ``snapshot'' approach of population synthesis, instead of the ``evolve'' approach. The snapshot approach is simpler but limited in its scope as it does not take into account any correlations between parameters. The evolve approach in particular is computationally expensive, and it was not feasible to run multi-parameter MCMC simulations with it.  Ideally, a faster implementation of population modeling can be used to constrain all the pulsar population parameters together. This would not limit the simulation by fixing some parameters, and would only require assumptions on the intrinsic distribution of the population parameters. Further, even with these constraints, the results from our analysis would greatly benefit from incorporating detection yields from more surveys, especially low ($<500$~MHz) and high ($>1.5$~GHz) frequency surveys. 

Also listed in Table 3 is the number of potentially observable MSPs (i.e.~those beaming towards us) in the Galaxy predicted by each model. These numbers are largely driven by our choice of luminosity function. For all models except C, we adopted
the log-normal parameterization obtained from the CP population \citep{2006faucher}. With that in mind, the range of potentially observable MSPs found here (14,000--39,000) are generally lower than found by 
\citet{2013MNRAS.434.1387L}. Further work in constraining the population size and beaming of MSPs in general is strongly encouraged.

\subsection{Predictions for current upcoming surveys}

We now use the parameters from our nominally optimal model (D) to make predictions for some ongoing (and upcoming) pulsar search surveys. These surveys are expected to find a lot of pulsars (both CPs and MSPs) and provide key insights into the population of radio pulsars. Table~\ref{tab:predictions} shows the parameters and number of MSPs detected with each survey. In addition to the emerging and future surveys listed in Table~\ref{tab:predictions},
PALFA \citep[see][and references therein]{parent2022} and the HTRU low-latitude \citep[][]{2020MNRAS.493.1063C} 
 surveys were not included as input to our analysis due to lack of accurate pointing information. Our model of the PALFA survey, which was only 71\% at the time of the collapse of Arecibo \citep{parent2022}, predicts 55$^{+8}_{-7}$ MSPs and agrees well with the number of MSP detections observed \citep[50, when previously known MSPs are included][]{parent2022} and is only slightly in tension with the HTRU low-latitude results \citep[26 MSPs;][]{2020MNRAS.493.1063C}.
 
 For all surveys except FAST GPPS, PALFA and HTRU, we have assumed an all-sky pulsar survey, i.e for the entire sky visible from the respective telescope. We have used nominal values for the integration time and other survey parameters. In many cases, such an all-sky pulsar survey would take several years to cover. Here, we do not account for the number of MSPs currently known. But these numbers are illustrative of the potential of these upcoming surveys to probe the Galactic MSP population.

\begin{table*}
\caption{Summary of {\tt PsrPopPy} runs showing the predicted sample sizes of MSPs ($N_{\rm MSP}$) detected by surveys planned in the upcoming decade. Also listed are ``postdictions'' for the PALFA and HTRU surveys (see text). }
\label{tab:predictions}
\centering
\begin{tabular}{lrrrrrr}
\hline 
Telescope & Band & Gain & Integration & Sky coverage &  $N_{\rm MSP}$ & References\\
& (GHz) & (K/Jy) & (s)  & ($^{\circ}$) &  & \\
\hline
{CHIME} & 0.4--0.8 & 2.0  & 900 & $\delta > -20 $  & $328^{+37}_{-30}$ & \citet{amiri2021}\\
{MeerKAT} & 1.1--1.8 & 2.0  & 600  & $-90 < \delta < +40$ & $464^{+45}_{-30}$ & \citet{bailes2020} \\
FAST  & 1.1--1.9 & 20 & 20  &  $-20 < \delta < +60$ & $451^{+44}_{-43}$ & \citet[][]{lorimer2019}\\
FAST GPPS & 0.8--1.7 & 16 & 300 & $-14 < \delta < +65$; $\vert b \vert <10$ & $482^{+36}_{-54}$ &  \citet{han2021} \\
ngVLA & 1.2--3.5 & 7.6 & 240 &  $-15<l<265$; $\vert b \vert <1$ & $84^{+12}_{-14}$ & \citet[][]{lorimer2019}\\
GBNCC & 0.3--0.4 & 2.0 & 120 & $-40 < \delta < +90$ & $94^{+6}_{-7}$ & \citet{stovall2014} \\
DSA2000 & 0.7--2.0 & 10 & 300 & $-40 < \delta < +90$; $\vert b \vert < 10$  & $792^{+87}_{-67}$ &  \citet{hallinan2019} \\ 
PALFA (71\% complete) & 1.2--1.5 & 8.2 & 268 & $32 < l < 77$; $\vert b \vert < 5$ & 55$^{+8}_{-7}$ & \citet{parent2022}\\ 
%PALFA outer & 1.2--1.5 & 8.2 & 180 & $168 < l < 214$; $\vert b \vert < 5$ & 1$^{+1}_{-1}$ & \citet[][]{parent2022, Lazarus2015}\\ 
HTRU Low (94\% complete) & 1.2--1.5 & 0.6 & 4300 & $-80 < l < 30$; $\vert b \vert < 3.5$ & 35$^{+7}_{-7}$ & \citet{2020MNRAS.493.1063C} \\ 
%HTRU Mid & 1.2--1.5 & 0.6 & 540 &  $-120 < l < 30$; $\vert b \vert < 15$& 46$^{+5}_{-7}$ & \citet[][]{keith2010} \\ 
%HTRU High & 1.2--1.5 & 0.6 & 270 & $\delta < +10$;  $15 <\vert b \vert < 90$   & 23$^{+4}_{-4}$ & \citet[][]{keith2010} \\ 

\hline 
\end{tabular}
\end{table*}

\subsection{Implications for MSP emission mechanism}

The analysis presented in this paper presents, for the first time, an empirical constraint on the population of MSP spectra at large. We have found evidence in favor of the population that deviates, on average, from a simple power law. In practice, there might be a number of factors at play which our analysis is not sensitive to. \citet{sieber1973} discussed synchrotron-self-absorption and thermal absorption to explain the low-frequency break observed in some pulsars. \citet{jankowski2018} attributed the observed deviation of pulsar spectral index from a simple power law to three scenarios: (1) environmental origin; (2) intrinsic spectral behaviour; (3) emission physics dependent on other pulsar properties (spin frequency, beam geometry, etc). In the first case, absorption due the local environment of MSPs leads to the observed features in an intrinsically featureless power-law spectral index. In the second case, the absorption processes originate in the magnetosphere of the pulsar, and so would be present throughout the pulsar population. \citet{meyers2017} observed a similar spectral break (and flattening) at low frequencies in giant pulses from the Crab pulsar, but it is uncertain whether this is related to the spectral break we report for the MSP population. A useful approach that could be applied to future studies which incorporate more MSPs as well as broader frequency constraints would be to investigate models in which a fraction of MSPs have simple power-law versus more complex spectra.

\section{Conclusions}
 
Using simulations to investigate the implications of the yields of large-scale pulsar surveys 
carried out a frequencies in the range $0.1<\nu<6.6$~GHz, we have presented a population analysis
of the spectral properties of millisecond pulsars, MSPs. The main conclusion of this work is that
a single power-law model cannot completely explain the population as it overestimates the number of MSPs found in low-frequency surveys. A far better description of the population is found when a two-component model is invoked, where flux density scales with frequency, $\nu$, roughly as $\nu^{-1.5}$ above about 320~MHz. Below this frequency, we find that flux density is approximately independent of frequency. The exact value of and behaviour of the spectrum below 300~MHz is currently uncertain, and further low-frequency surveys and follow-up studies of individual MSPs will be insightful in this area. Our optimal model predicts a substantial population of MSPs to be discovered from current and upcoming surveys in the next decade.

For the Galactic MSP population in general, the current sample is now\footnote{For an up-to-date list of MSPs currently known in the Milky Way, see http://astro.phys.wvu.edu/GalacticMSPs.} in excess of 400. By comparison, this now exceeds
the set of canonical pulsars (CP) available to \citet{1985MNRAS.213..613L} in their classic study of the
CP population. Looking back on the CP population literature over the past four decades, a wide variety of issues including magnetic field evolution \citep[e.g.,][]{1992A&A...254..198B} and initial spins
\citep[][]{1993MNRAS.263..403L} have been explored. Similarly with MSPs, we anticipate a number of
insightful studies which make use of the existing population to further constrain their properties.

\section*{Acknowledgements}
K.A. acknowledges support from National Science Foundation (NSF) grant AAG-1714897. We thank Kaustubh Rajwade, Devansh Agarwal, Paul Demorest, Tyler Cohen and Maura McLaughlin for useful discussions. We also thank Alex McEwen, Kevin Stovall, Julia Deneva and Joe Swiggum for providing details of AODRIFT and GBNCC surveys.  We acknowledge use of the Spruce Knob supercomputer at WVU, which are funded in part by the NSF EPSCoR Research Infrastructure Improvement Cooperative Agreement \#1003907, the state of West Virginia (WVEPSCoR via the Higher Education Policy Commission) and WVU.

\bibliographystyle{mnras}
\bibliography{msp}

% Don't change these lines
\bsp	% typesetting comment
\label{lastpage}
\end{document}